\documentclass{iau-JDSS}
\usepackage{graphicx}

\title[short title of paper] %% give here short title %%
{Galaxy formation from dry and hydro simulations}

\author[short author list]   %% give here short author list %%
{Luca Ciotti}%
\affiliation{Department of Astronomy, University of Bologna,
via Ranzani 1, 40127 Bologna, Italy}
\pubyear{2009}
\volume{Volume 15}  %% insert here IAU Highlights of Astronomy Volume Number
\pagerange{1}
\date{?? and in revised form ??}
\jname{Highlights of Astronomy, Volume 15}
\editors{Ian F Corbett, ed.}
\begin{document}

\maketitle

\begin{abstract}
  The effects of dry and wet merging on the Scaling Laws (SLs) of
  elliptical galaxies (Es) are discussed.  It is found that the galaxy
  SLs, possibly established at high redshift by the fast collapse of
  gas-rich and clumpy stellar distributions in preexisting dark matter
  halos following the cosmological SLs, are compatible with a
  (small) number of galaxy mergers at lower redshift.
\end{abstract}

%\section{Introduction}

The main results obtained in a series of papers (Ciotti \& van Albada
2001; Nipoti, Londrillo \& Ciotti 2003; Lanzoni \etal\ 2004; Ciotti,
Lanzoni \& Volonteri 2007; see also Ciotti 2009), are presented.  It
is found that 1) Parabolic dry merging in a population of low mass
spheroids leads to massive Es that fail the Faber-Jackson (FJ) and
Kormendy relations, being characterized by low velocity dispersion and
very large effective radii. Parabolic wet merging in the same
population of progenitors leads to Es in better agreement with the
observed SLs, as long as enough gas for dissipation is available.  2)
The edge-on structure of the Fundamental Plane (FP) is surprisingly
preserved. Therefore, the FJ and Kormendy relations, despite their
larger scatter, are stronger tests for merging than the edge-on FP.
3) Parabolic dry or wet merging of Es following the observed SLs over
the full mass range preserve the Kormendy, FJ, and edge-on FP
relations.  Thus, massive Es cannot be formed by parabolic merging of
low mass spheroidal galaxies, even in presence of substantial gas
dissipation, but their SLs, once established by galaxy formation, are
robust against merging.  4) Dark matter halos obtained from
cosmological simulations define a FJ, a Kormendy, and a FP-like
relation, as expected from the spherical collapse model for virialized
systems.  5) Numerical simulations of cold dissipationless collapse in
pre-existing dark matter halos can reproduce Sersic profiles
remarkably similar to those observed, over a large radial range. Note
that cold dissipationless collapse is a process which is expected to
dominate the late stages of an initially dissipative process. Thus the
SLs of Es, possibly established at high redshift by the fast collapse
of gas rich and clumpy stellar distributions in pre-existing dark
matter halos (following the cosmological SLs), can persist even in the
presence of a moderate number of dry or wet mergings.  Then
monolithic-like collapse at early times and subsequent merging could
just represent the different phases of galaxy formation (collapse) and
evolution (merging, in addition to the aging of the stellar population
and related phenomena).

\end{document}